# Combined Oxy-fuel Magnetohydrodynamic Power Cycle

Osama A. Marzouk

*University of Buraimi, college of engineering, Mechanical department*
*Al Buraimi, Sultanate of Oman*
osama.m@uob.edu.om

**Abstract**

Oxy-fuel carbon capture in power plants is a relatively new concept aiming at reducing carbon dioxide emissions from the plants. This is achieved by burning the fossil fuel using oxygen as oxidizer with no nitrogen, thereby rendering the exhaust gases very rich in carbon dioxide (after condensing water vapor by cooling), which facilitates its capture for environmental or commercial purposes. Despite the worldwide interest in oxy-fuel carbon capture, its progress is at risk given the large energy needed to separate oxygen from air in order to provide the oxidizer, thereby hindering further progress of this concept toward large-scale applications. This paper focuses on alleviating this drawback of oxy-fuel combustion by making it more attractive through combining it with another concept, namely magnetohydrodynamic (MHD) power generators. The end product is a power plant operating on a combined cycle composed of a topping MHD ultra-high-temperature cycle with direct electricity extraction from plasma, followed by a bottoming steam cycle with conventional turbo-generators. Different design aspects and simplified technical analysis for the MHD generator are presented.

**Introduction**

Magnetohydrodynamic power generators (Messerle, 1995; Kayukawa, 2004; Woodside et al., 2012) are chambers where high-speed electrically-conducting hot plasma expands while subject to an applied magnetic field. This leads to the generation of direct-current electricity based on Faraday's law of induction without the conventional turbo-generators. MHD requires ultra-high temperatures (near 3000 K) for attaining reasonably-high levels of electrical conductivity of the plasma.

Oxy-fuel combustion (Wall et al., 2009; Zheng and Tan, 2014) is another concept where fossil fuels are burnt using a nitrogen-free oxidizer, leading to ultra-high temperature gases which can be made rich in $CO_2$ after condensing their $H_2O$ content through cooling. The $CO_2$-rich flue gas in oxy-fuel combustion facilitates $CO_2$ capture (hence the term 'carbon capture') and then sequestering it in underground formations instead of releasing it into the atmosphere, in an attempt to mitigate the massive emissions of this primary greenhouse gas from anthropogenic activities (Liu et al., 2012). Moreover, the captured $CO_2$ can be sold commercially for later use in enhanced oil recovery by injecting supercritical dense-phase $CO_2$ into the oilfield to reduce the viscosity of the crude oil and enhance its extraction.

The ultra-high temperature is thus a common feature between the MHD concept and the oxy-fuel concept, raising the question of combining them if a benefit can be achieved as a consequence. A main drawback of oxy-fuel combustion as a concept to be used in steam power plants is the large energy penalty incurred by separating oxygen from air. This problem is so important that the giant Swedish energy company Vattenfall, which previously announced long-term plans for research investments in oxy-fuel combustion at demonstration scale and commercial scale, has decided in May 2014 to stop all effort on oxy-fuel carbon capture due to "its costs and the energy it requires makes the technology unviable". The company gave priority to other R&D projects "which can contribute more quickly to our business development". This decision aborted an already ongoing project for building a 250 MWe demonstration plant in Jänschwalde (Germany) to be commissioned in 2016, despite the successful commissioning of a smaller oxy-fuel demonstration plant (to produce steam for nearby industry rather than electricity) with 30 MWth capacity at the Schwarze Pumpe power station (Germany) since September 2008 (Anheden et al., 2011). On the other hand, a main drawback of the MHD concept as a potential concept for power generation is that it exploits energy only in the very high temperature spectrum, leaving hot gases with huge amounts of unexploited energy. However, having a very high working-fluid temperature in MHD generators enables extremely high levels of energy extraction from this working-fluid (the plasma), which mitigates the energy consumption in air separation units (ASU) for oxy-fuel combustion. In the same time, steam cycle can be used to extract energy from the still-hot MHD exhaust gases, where these gases operate the steam generator to provide the necessary steam for the steam turbine without using a furnace. Thus, the steam cycle handles the extraction of the energy from the working-fluid at the lower temperature spectrum (about 2500 K and below). This releases the MHD limitations and makes it a viable concept for commercial power plants, and in the same time makes oxy-fuel combustion more attractive with the mitigated effect of the air separation cost.

**What is a Magnetohydrodynamic Generator?**

Figure 1 illustrates a typical design of an MHD generator. It also shows the Fleming's right-hand rule, governing the direction of the induced current due to the motion of the conducting plasma within the applied magnetic field. The generator consists of

1) Combustor:
   - The fuel is burnt with $O_2$, plasma is generated.

2) Convergent passage:
   - Plasma accelerates to sonic speed at the throat.

3) Channel:
   - This is the heart of the MHD generator.
   - Plasma continues to accelerate with supersonic speeds.
   - Electrodes (anode and cathode) are fitted to two opposite walls.
   - An external magnetic field is applied in the bi-normal direction (normal to both the electrodes and the bulk motion of the plasma).
   - Energy is extracted from plasma; electric current is collected through the electrodes.

4) Convergent diffuser (not shown):
   - To slow down the exploited plasma.

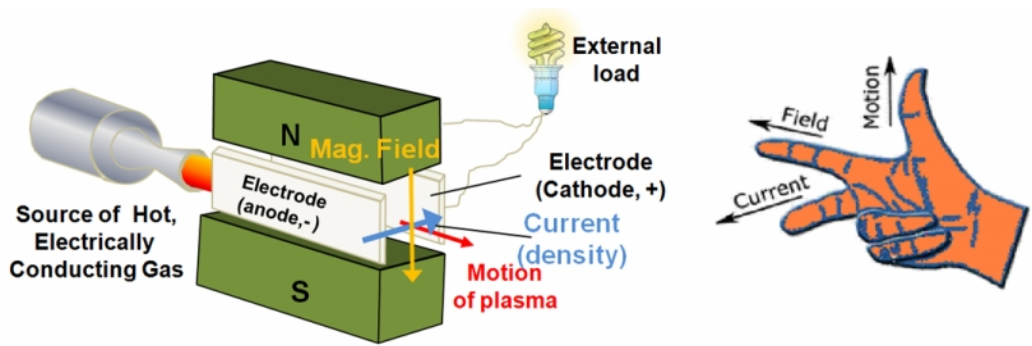

**Figure 1. Ilustration of MHD Generator.**

While the use of a divergent channel with supersonic plasma is not strictly mandated, it is highly demanded because the power output from the MHD generator is proportional to the square of the plasma velocity. In addition, as energy is extracted from the plasma, it cools down near the rear section of the MHD channel, thereby dropping its electrical conductivity in a sharp manner. This should be compensated by an increase in the plasma velocity, which is again achieved by a supersonic divergent channel where the plasma continues to accelerate downstream the channel despite the loss of enthalpic energy.

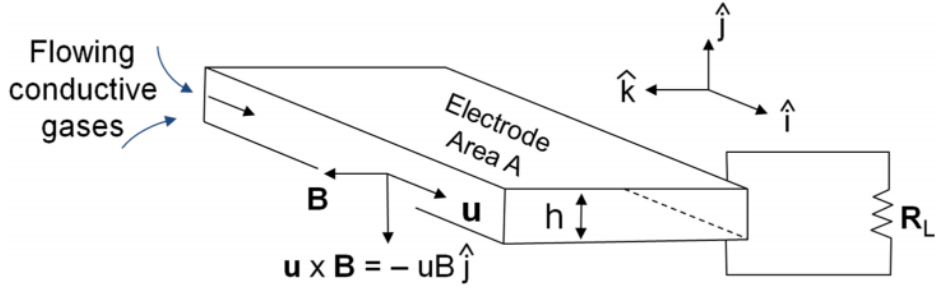

**Figure 2. Simple 1-D MHD Channel.**

Figure 2 further demonstrates the working principle of a simplified one-dimensional MHD channel, having a constant cross-sectional area with a fixed spacing ($h$) between parallel electrodes. The plasma moves in the axial direction (x-axis) with a velocity ($\vec{u} = u\,\hat{\imath}$), the external magnetic field ($\vec{B} = B\,\hat{k}$) is applied in the z-direction, and the induced electric field ($\vec{u} \times \vec{B} = -u\,B\,\hat{\jmath}$) is in the negative y-axis, going from the anode (upper, negative electrode) to the cathode (lower, positive electrode). The external load is represented by a resistance ($R_L$) connected externally to the electrodes.

## Plasma Formation for MHD Generator

While the hot gases coming from the MHD combustor posses very high temperatures (e.g., 3200 K) that enable partial ionization, the electrical conductivity of these gases (mainly $H_2O$ and $CO_2$) will still be negligible because of the relatively-large ionization energy for their molecules. To remedy this, combustion products are augmented with the vapor of an alkali metal (such as cesium: Cs, or potassium: K) at a small mole fraction of about 1%. The alkali metal can be introduced into the combustor in a form of seeded salt compound, such as potassium carbonate ($K_2CO_3$). It should be regenerated and re-used for economically-feasible operation. In fact, cesium has the lowest ionization energy among elements, meaning that it is the easiest element to lose an electron and form an ion. On the other hand, helium (He) is the element having the largest ionization energy. Table 1 compares the ionization energies of cesium, potassium, hydrogen, oxygen, and helium (Kramida et al., 2014). Whereas cesium has lower ionization energy than potassium, it is much rarer.

| element | Ionization energy (eV/atom) | Ionization energy (kJ/mol) |
|---|---|---|
| Cesium (Cs) | 3.89 | 375 |
| Potassium (K) | 4.34 | 418 |
| Hydrogen (H) or Oxygen (O) | 13.6 | 1310 |
| Helium (He) | 24.6 | 2370 |

**Table 1. Comparison of the Ionization Energy for Selected Elements.**

Figure 3 (Swithenbank, 1974) emphasizes the strong dependence of the electrical conductivity of the alkali-seeded plasma on the temperature. The data are for oxy-fired JP-4 combustion products under 1% potassium seed. JP-4 (Jet Propellant 4) is a military aviation fuel having 50-50 kerosene-gasoline composition by volume. For alkali-metal ionization in equilibrium plasma, Saha equation (Saha, 1920) predicts that the number density of released electrons ($n_e$) is strongly affected by the temperature, obeying the following the proportionality expression:

$$n_e \propto T^{1.5} \exp\left(-\frac{IE}{\hat{k}_B T}\right) \tag{1}$$

where $T$ is the temperature (in kelvins), $IE$ is the ionization energy (in eV, electron volts), $\hat{k}_B$ is Boltzmann constant expressed in eV/K (having the constant value of $8.617 \times 10^{-5}$).

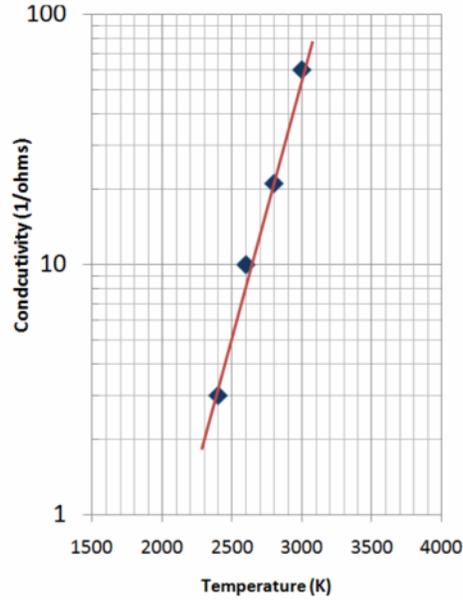

**Figure 3. Sensitivity of Electrical Conductivity to Temperature (Oxy-fired JP-4 seeded with 1% potassium).**

## Simplified Electrical Analysis of MHD Channel

A simple analysis is presented here to estimate the maximum electrical power output from an MHD channel. We use in Figure 4 the simplified 1-D channel of Figure 2, which greatly facilitates the discussion and permits analytical expressions.

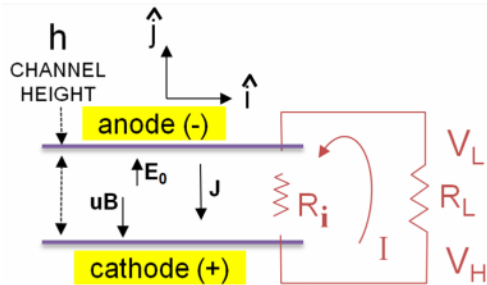

**Figure 4. Electric Analysis of a Simple 1-D MHD Channel**

The current flux (current density vector) is denoted by $J$, which points to the negative y-axis. The external electrical field applied in the channel is

$$E_0 = \frac{V_H - V_L}{h} \qquad (2)$$

where $V_L$ is the low voltage (at anode) and $V_H$ is the high voltage (at cathode), both are relative to the ground. The applied magnetic field $B$ (in teslas) is in the z-axis, coming out of the page. There is an internal resistance formed by the plasma inside the channel and is represented by $R_i$, where

$$R_i = \frac{h}{\sigma A} \qquad (3)$$

with $\sigma$ being the electrical conductivity of the plasma, and $A$ is the projected channel area in the x-z plane, which is also the area of either electrode.

The current *I* in the external circuit is related to the current density across the plasma by $I = J\,A$. The magnitude of current density is calculated as

$$J = \sigma\,(u\,B - E_0) \qquad (4)$$

The maximum voltage across the electrodes is the open-circuit voltage $V_{O.C.}$, occurring when $I = J = 0$ ($R_L = \infty$). In this case, manipulating (4) and (2) gives that this voltage is:

$$V_{O.C.} = u\,B\,h \qquad (5)$$

In the normal operating case, $J \neq 0$ ($R_L$ is finite), and the current is

$$I = \frac{V_{O.C.}}{\sum R} = \frac{u\,B\,h}{R_L + R_i} = \frac{u\,B\,h}{R_L + \dfrac{h}{\sigma\,A}} \qquad (6)$$

The power to the load is $P_L = I\,V_L = I^2\,R_L$, where $V_L$ is the voltage across the load. In the case of an open circuit; $R_L = \infty$, $I = 0$, and $P_L = 0$. In the case of a shorted load; $R_L = 0$, $I = u\,B\,h/R_i = u\,B\,\sigma\,A$, and again $P_L = 0$. Introducing a load factor K $= R_L/\sum R = R_L/(R_L + R_i) = E_0/(u\,B)$, and then expressing $P_L$ in terms of that factor, it can be shown that between these two extreme conditions $P_L$ has a maximum occurring at the condition of K = ½, or $R_L = R_i$. This maximum value of the load power is

$$P_{L,\max} = \frac{\forall\,\sigma\,u^2\,B^2}{4} \qquad (7)$$

where $\forall$ is the channel volume, $A\,h$.

Considering some representative values for a proposed MHD channel, we have:

$\forall = 6 \times 2 \times 1 = 12$ m$^3$, $\sigma = 15$ S/m (siemens/m), $u = 1200$ m/s, $B = 6$ T (tesla). This predicts the maximum electric power to the load to be 2333 MWe = 2.3 GWe. This corresponds to a paramount power density (power per volume) of 194 MWe/m$^3$.

Although this estimate is rough because it is based on simplified analysis, it is quite conservative because the magnetic field can have larger values with the presence of today's super magnets. Such magnets are already used in the medical field as magnetic resonance imaging (MRI) equipment; and in scientific research as nuclear magnetic resonance (NMR) spectrometers, mass spectrometers, and particle accelerators. They are electromagnets but constructed using coils of a superconducting wire. Thus, they must be cooled to cryogenic temperatures to attain the superconducting phase, where the superconducting wire conducts much larger electric currents than an ordinary wire due to diminished resistance, thereby creating intense magnetic fields. Superconducting magnets can outperform fields all but the strongest ordinary electromagnets in terms of the produced magnetic fields and also can be cheaper to operate because no energy is dissipated as heat in the windings. For the plasma velocity, if the plasma gases have a gas constant of $R = 235$ J/kg-K (say 33% $H_2O$ - 67% $CO_2$, by volume), and have a ratio of specific heats ($\gamma$) of 1.15, and an average temperature of $T = 2850$ K, then the speed of sound within the plasma is $a = \sqrt{\gamma\,R\,T} = 878$ m/s. The above taken value thus corresponds to a modest Mach number of about 1.4. Even if the actual power is 50% of the above estimate, the power density remains attractive.

## Layout for Combined Oxy-fuel MHD Cycle

Figure 5 shows a schematic layout for a power plant adopting the combined oxy-fuel MHD cycle, described here. The inverter is needed to convert the DC electric output to AC output, suitable for the power grid. The seed regenerator is responsible of recycling the seeded compound of alkali metal back to the combustor.

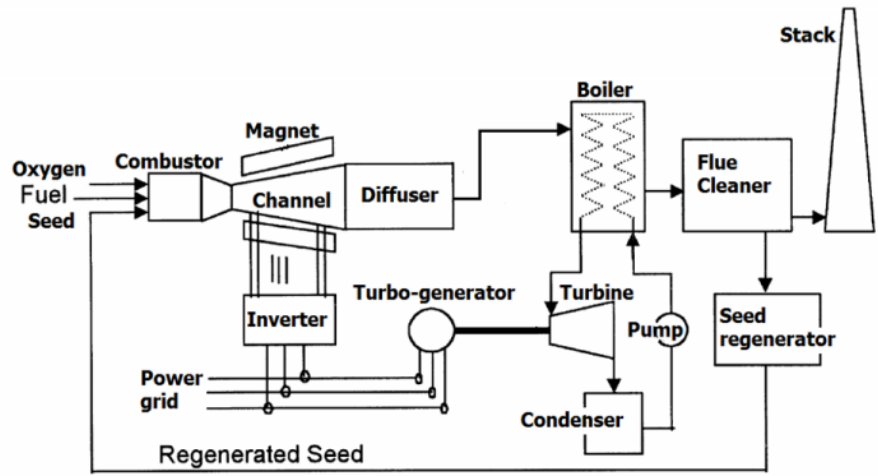

**Figure 5. Layout of a Power Plant Utilizing the Combined MHD Oxy-fuel Cycle.**

## Conclusions

A combined oxy-fuel magnetohydrodynamic (MHD) power cycle was proposed and discussed as a means to counteract drawbacks of the separate concepts of oxy-fuel and MHD generator by capitalizing on the advantages of each. The cycle combines a topping MHD generator with a bottoming steam plant. Some design aspects were presented and a simplified 1-D model predicts that more than 2 GWe output can be produced from a compact space of only 12 m$^3$.

## Acknowledgements

The author has benefited greatly from different resources made available to him at the National Energy Technology Laboratory or the U.S. Department of Energy. The author is especially grateful to E. David Huckaby and Geo Richards.